\documentclass[final,5p,times,twocolumn,authoryear]{elsarticle}

\usepackage{lineno,hyperref}
\usepackage{graphicx}

\journal{Astronomy and Computing}

\biboptions{authoryear}

\begin{document}

\begin{frontmatter}

\title{JOVIAL: Notebook-based Astronomical Data Analysis in the Cloud}

\author[i1]{Mauricio Araya\corref{mycorrespondingauthor}}
\author[i1]{Maximiliano Osorio}
\author[i1]{Mat\'ias D\'iaz}
\author[i1]{Carlos Ponce}
\author[i1]{Mart\'in Villanueva}
\author[i1]{Camilo Valenzuela}
\author[i1]{Mauricio Solar}
\address[i1]{Universidad T\'ecnica Federico Santa Mar\'ia, Avenida Espa\~na 1680, Valpara\'iso, Chile}

\cortext[mycorrespondingauthor]{Corresponding author e-mail: \texttt{mauricio.araya@usm.cl}}




\begin{abstract}
Performing astronomical data analysis using only personal computers is becoming impractical for the very large data sets produced nowadays. As analysis is not a task that can be automatized to its full extent, the idea of moving processing where the data is located means also moving the whole scientific process towards the archives and data centers.
Using Jupyter Notebooks as a remote service is a recent trend in data analysis that aims to deal with this problem, but harnessing the infrastructure to serve the astronomer without increasing the complexity of the service is a challenge.
In this paper we present the architecture and features of JOVIAL, a Cloud service where astronomers can safely use Jupyter notebooks over a personal space designed for high-performance processing under the high-availability principle. We show that features existing only in specific packages can be adapted to run in the notebooks, and that algorithms can be adapted to run across the data center without necessarily redesigning them.
\end{abstract}


\end{frontmatter}


\section{Introduction}

The fast-paced technology improvements on astronomical telescopes, instruments and archives,  
plus the accumulative nature of astronomical data, 
contribute to generate an overwhelming data space waiting to be analyzed and exploited.
The data deluge problem in astronomy is not a menacing challenge in the
horizon, but a very real challenge for current astronomical data analysis.
This is the case for the ALMA Data, 
where some single data products can reach hundreds of GigaBytes, being expensive to 
transport, store, process, and load them into memory \citep{testi10}. 
The next generation of projects such as Large Synoptic Survey Telescope - LSST \citep{ivezic2008lsst}, Extreme Large Telescope - ELT \citep{gilmozzi2007european},  and the Square Kilometer Array - SKA \citep{dewdney2009square} 
will increase the data generation rate from two to three orders of magnitude,
so scientists and engineers are preparing themselves to cope with this new reality.

The somewhat mundane problem of personal computers not being able to store and process the astronomical files produced nowadays, have been forcing a paradigm-shift in astronomical data analysis. It is not enough to develop new algorithms and allocate computing infrastructure; modern software and services need to be fostered to equip astronomers with the tools to deal with the data deluge problem.

It is clear that processing tasks need to be moved towards the large archives where the data resides, minimizing the data transfer to the end user and taking advantage of the high-performance computing infrastructure available at the data centers \citep{djorgovski2003challenges,moolekamp2015toyz}.
However, the efficient use of the high-performance computing and storage infrastructure requires a good understanding of technical details that are not in the domain of expertise of astronomers. Moreover, new challenges arise when these services are deployed under the high-availability principle in order to compete with the convenience of local processing. 

This paper presents our approach to provide a Cloud service for astronomical data analysis called JOVIAL, which provides Jupyter notebooks to astronomers in a transparent and high-available fashion. Section~\ref{sec:nada} discuss why Jupyter notebooks are a suitable tool for astronomical data analysis, while Section~\ref{sec:jovial} presents the architecture that we used for offering them in the Cloud. Then, Sections \ref{sec:toolkits} and \ref{sec:distributed} presents our current efforts to provide a consistent set of libraries for astronomical data analysis and how the computing tasks can be distributed along the available infrastructure. We conclude in Section \ref{sec:conclusions} with our next steps towards improving the services provided by JOVIAL.

\section{Notebook-based Astronomical Data Analysis}
\label{sec:nada}

Modern data analysis strongly relies on computational methods, models and algorithms.
These are usually grouped in software packages that provide from very general to very specific functionality that support the analysis process. Despite the large variety of them, one can identify the key elements that these set of tools must cover. 

\paragraph{Interactivity} 
Most of the visualizations and plots that are made during the analysis process are not published nor shared, and only the key results are published. Any discovery process include some trial and error actions, fine tuning and data exploration tasks. Interactive tools help to speedup this discovery process, and that is the main reason why astronomers usually use specialized interactive desktop applications. 

\paragraph{Documentation}
All the findings of an analysis process need to be documented somewhere: the parameters used in the applications, the data transformations, the formulas and insightful comments need to be registered in some document. Usually, interactive desktop applications provide information in the form of action history and logs, but these are in a suitable format for being reported. The documentation process need to be carried out under a different tool such a document preparation software.

\paragraph{Automation}
When a task needs to be executed repeatedly for a collection of data or for different parameters (i.e., batch processing), the automation is carried out using high-level programs called scripts.
The documentation of these procedures are usually included in the same scripts as comments.
Depending on the tools used for analysis, the execution of these scripts can be supported by the same analysis tools, or they need to be loaded into a different system. A typical example of the latter is when a time-consuming pipeline need to be executed in a computer cluster.

\subsection{Interactive Notebooks}

Interactive notebooks are files containing both code and documentation that can be interactively edited, displayed and executed \citep{shen2014interactive}. 
The name comes from the analogy to paper notebooks, where ideas can be drafted in a dynamical way.
Notebooks were first adopted by mathematical software packages like Maple, SageMath and Mathematica, and they have been used to assist researchers in the scientific workflow for all sciences since then \citep{kluyver2016jupyter}. Most notably, major scientific breakthroughs like the first gravitational wave detection \citep{abbott2016observation}, have been accompanied by notebooks to share the exact data analysis process that was followed. 

The notebook concept consist then in combining \emph{interactivity}, \emph{documentation} and \emph{automation} in a single file, and therefore, using a single comprehensive tool for data analysis. The interactivity allows to understand a problem by performing small, incremental improvements that lead to a complete analysis. The documentation allows sharing, from early to publication stages, a valuable computational narrative that it is difficult to present otherwise. The automation allows to generate self-documented, exportable and reproducible pipelines with graphical support.

\subsection{Jupyter Notebooks}

Jupyter is an open-source web application for authoring interactive notebooks that has gained an important popularity in recent years. For example, in January of 2018 the number of  Jupyter notebook files in Github was over 1.6 millions\citep{parente2018nbestimate}.
The Jupyter project is the evolution of the IPython interactive shell, which since 2001 has provided tools to extend Python’s interactive capabilities \citep{perez2007ipython}.
Currently, Jupyter support several \emph{kernels} types, which are the interpreter processes that actually execute the inline code inside the notebooks. This feature of Jupyter extends its support to several programming languages, including compiled ones.
Besides authoring notebooks, Jupyter web interface includes simple process and file managers, an spawning terminals system  and extensions that allow to graphically manage packages and parallelism. The JupyterLab project \citep{jlab2018jupyterlab}  extends these in-browser capabilities to produce a window-manager-like interface, including text editors, documentation viewers, resource monitors, etc.   

\begin{figure}[h]
\centering\includegraphics[width=0.8\linewidth]{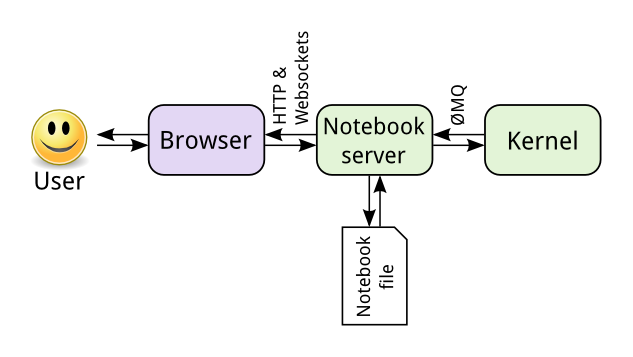}

  \caption[Jupyter Notebook Architecture.]{Jupyter Notebook Architecture. Figure obtained from \url{http://jupyter.readthedocs.io}}
\label{juparch}
\end{figure}
The architecture of Jupyter notebooks shown in Figure \ref{juparch}, detaches the presentation logic from the business logic, being the first one handled by the browser through Javascript code, and the second one by a service written in Python. This allows Jupyter to run in a web-based client-server model. At the server, the architecture also detaches the processes in charge of interpreting the notebooks files from those in charge of interpreting the code snippets in the notebook (i.e., the kernel). This allows kernels to halt or be restarted without compromising the service itself, and also it ease the support of kernels outside Python. The communication between these processes is done with an efficient JSON messaging using the ZeroMQ library.

This architecture makes it possible to use Jupyter as a remote notebook authoring application that executes arbitrary code within the host server. This opens up several opportunities and challenges that we address in Sections~\ref{sec:jovial}, \ref{sec:toolkits} and \ref{sec:distributed}.

\subsection{Notebooks and Astronomical Libraries}

There are several software packages for performing astronomical data analysis; from simple file viewers to complete pipeline toolkits. Each package have its own scope, semantics and installation dependencies. 
Simple operations may overlap between packages,
but there are others that are provided only by one specific package. 
Currently, an astronomer must install all the software tools that he might need, and make all the data and metadata conversions needed to inter-operate between these tools.
Some efforts of standardization and inter-application communication have been developed by the virtual observatory community, yet not all packages support protocols like SAMP nor publish data as VoTables \citep{taylor2012samp}. 

While Jupyter supports languages that traditionally have been used for astronomical data analysis like R, Matlab and IDL, most of the available notebooks for astronomy use the IPython kernel. This is for two major reasons: Before all these languages were supported by Jupyter, IPython notebooks were already very popular \citep{shen2014interactive}, and Python itself is a major trend in astronomy \citep{greenfield2011python}. 

There are many Python libraries focused in generic scientific data analysis, such as 
\texttt{NumPy},
\texttt{SciPy},
\texttt{Pandas},
or
\texttt{Matplotlib}. These are actually \emph{notebook friendly}, which means that their outputs are compatible or have an special extension to be better displayed into notebooks. There are also several astronomical data analysis tools written in Python, but not all of them are designed for execution within interactive notebooks. The \texttt{astropy} project \citep{robitaille2013astropy}, and the important number of affiliated packages built over it, have used notebook-friendly packages and developed the necessary features for notebook compatibility. 
On the other hand, projects like CASA \citep{mcmullin2007casa}, which actually uses the IPython console, does not offer a compatible interface to notebooks. This is because its graphic user interfaces are designed as stand-alone applications in QT. 



\subsection{Jupyter as a Cloud Service}

Using Jupyter as a remote service is a recent trend within astronomical data centers. Several institutions are deploying their own Jupyter-based architecture to support astronomers in their work. For instance, SciServer Compute \citep{medvedev2016sciserver} uses Jupyter notebooks attached to large relational databases and storage space to bring advanced data analysis capabilities closer to the data. 
Another example is the NOAO Data Lab \citep{fitzpatrick2014noao}, which provides a collaborative computing environment where users can easily access and visualize large datasets, and conduct experiments on subsets of the data using Jupyter notebooks.

Offering notebooks as a web-based service is the first step towards moving interactive analysis from the user' computers to where the data resides (i.e., data centers). However, several gaps must be filled in order to produce a quality service that astronomers can trust and rely on. A Cloud service is not only a web-application, but a service that hides the complexities of the deployment, maintenance and high-availability configurations of dynamic resources over one or several data-centers.

From the user point of view, notebooks running in the Cloud offer several benefits such as world-wide access, non-blocking computations, installation-free experience, augmented storage and computing resources, task distribution, and collaboration opportunities. 
However, each one of those imposes new challenges, such as difficulties to work off-line, synchronization problems, custom installation of libraries, personal backups, managing distributed resources, and privacy issues.

In the following section, we describe our approach to deal with some of these challenges.


\section{JOVIAL: Notebooks in the Cloud}
\label{sec:jovial}

\begin{figure}[h]
\centering\includegraphics[width=0.95\linewidth]{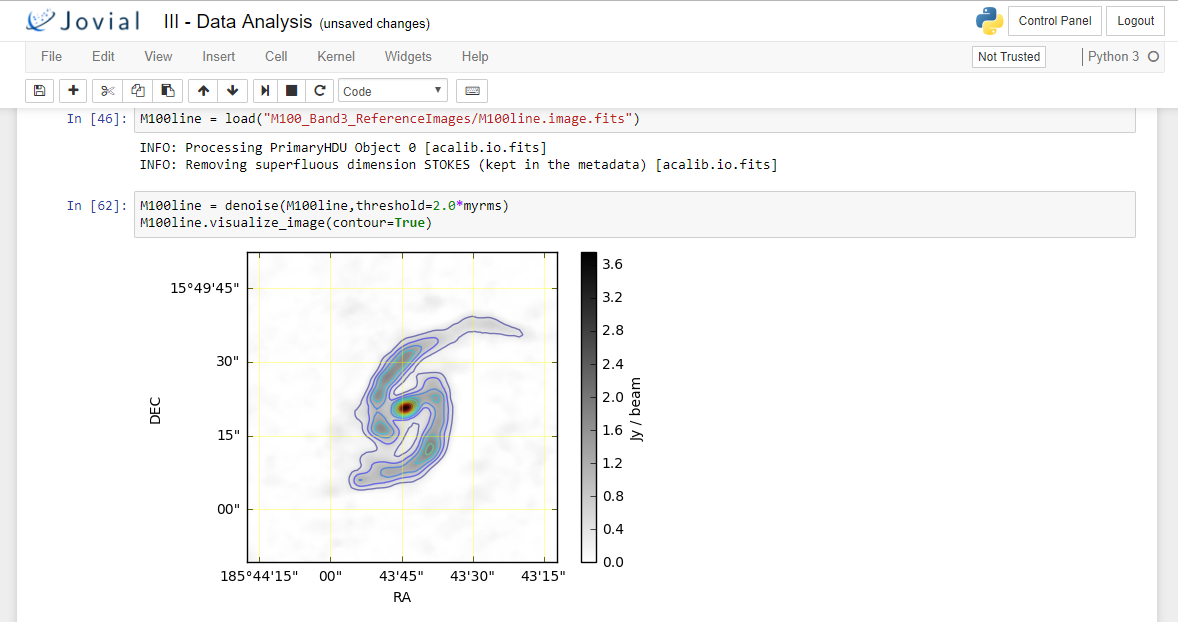}
\caption{Jupyter Notebook Running on JOVIAL.}
\label{fig:jovial_example}
\end{figure}

JOVIAL\footnote{Jupyter OVerride for Astronomical Libraries.} is a Cloud service developed by the Chilean Virtual Observatory \citep{solar2015chilean}, where astronomers can author their own Jupyter notebooks over a personal space designed for high-performance processing \citep{araya16adass}. 
The main goals of JOVIAL is to bring data analysis closer to the data, take advantage of high-performance infrastructure, and protect the users' data and results. An screenshot of JOVIAL running is shown in Figure~\ref{fig:jovial_example}.

The architecture of JOVIAL \citep{diaz17adass} is composed by several software, systems and hardware components that we have incrementally included to provide a better service. In this section we describe and explain how we use these components, in order to be able to replicate our configurations. In the last part of this section we present the complete architecture.

\subsection{Multi-User Jupyter Service}

The original Jupyter notebook server allows only one user per instance, meaning that each interface need to be bound to a different port. JupyterHub is a multiple-user version of Jupyter, where each user owns its own server but through the same instance. The process of starting a Jupyter notebook server for a user is known as \textit{``spawn''}, and where it spawns is determined by JupyterHub modules called spawners \cite{jhub2018jupyterhub}.

The concept that each user spawns a server might seem unnecessary at first glance, but remember that notebooks are just one part of the Jupyter interface. Users need to manage their files, install custom packages, manage the running kernels, etc. All these operations must run with an user ID for accountability and access control. 

\subsection{Docker Containers}

The main problem with JupyterHub is that by default the user of the account is a generic system user, and the notebook spawns in the host machine. This combined with the feature of running arbitrary code in the notebooks represents a risk for the data center and for the other users.


To solve this problem we decided to spawn the services in isolated environments, where users are restricted 
to use only the resources that are assigned to them. Also, this isolation provides a security layer, since users are not even aware of the existence of other users. Isolation of the notebooks can be achieved through virtualization, for which we found two approaches: the traditional hypervisor-based virtualization and the relatively new container-based virtualization. The first one provides virtualization at the hardware level creating and running virtual machines. Vmware \citep{walters1999vmware}, Xen \citep{barham2003xen} , and KVM \citep{kivity2007kvm} are examples of Hypervisor-based virtualization. The second one is a lightweight virtualization approach that uses the host kernel to run multiple virtual environments. This provides isolated environments with only the necessary resources to run the applications \citep{bui2015analysis}.

We have chosen container-based virtualization because studies show that containers have a performance close to a native application \citep{xavier2014performance} and they are fairly secure \citep{bui2015analysis}. We use the Docker system \citep{merkel2014docker} , which is an open-source container technology with the ability to build, ship, and run distributed applications. In our context, containers recreate a system environment inside the host, where we can spawn a notebook server safely, including the personal user files through Docker volumes. Similar projects to JOVIAL, like SciServer Compute \citep{medvedev2016sciserver}, are also using Docker containers for the same purpose. 

To control the resources used by the users, we use a Linux kernel feature called control groups (cgroups). A cgroup limits an application to an specific set of resources. Cgroups allows Docker to enforce limits and constraints, so it is possible to limit the memory, CPU, disk, and others to an specific container and user \citep{higgins2015orchestrating}. 

\subsection{Orchestration Over Multiple Hosts}

\begin{figure}[h]
\centering\includegraphics[width=0.8\linewidth]{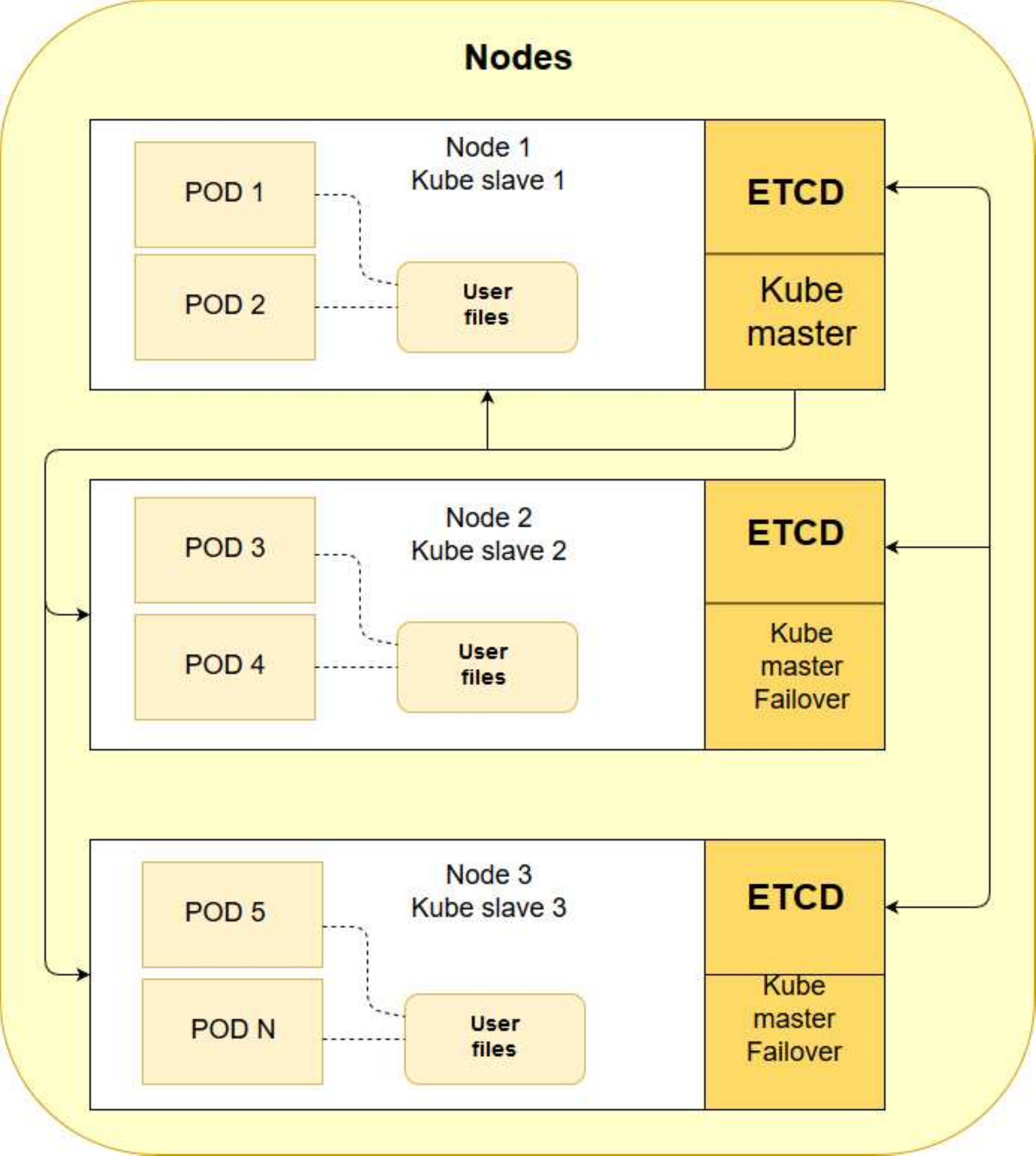}
\caption{Kubernetes Deploy, each host has a database named etcd. One node is the master if one node fails the others keep Kubernetes working, if the master fails, the other ones vote for a new master and keep working.}
\label{kube1}
\end{figure}

The notebook servers can run on a single host or multiple hosts. By keeping them in a single host, we are taking the risk of centralizing the infrastructure, so the failure of this single host will compromise the whole platform. 
Also, the multiple-hosts approach allows us also to use better our data-center resources. But containers deployed within multiple hosts needs to be orchestrated in accordance with the high-availability principle, meaning that they should be robust and self-healed under standard hardware or network malfunctioning.



In order to achieve this, we combined our Docker containers with Kubernetes.
Kubernetes is an open-source system for automating deployment, scaling, and management of containerized applications called PODs \citep{burns2016borg}. 
Kubernetes decides where to run the container based on the current load of the nodes and permissions, providing on-demand growth and load balancing between the hosts. 
For networking, Kubernetes uses a service called Flanneld that gives each container an IP address in a logical private network (different from the one of the nodes) that connects the containers across the nodes \citep{marmol2015networking}. 

If one node fails, Kubernetes will migrate all the containers to other available nodes \citep{bernstein2014containers}. 
So Kubernetes is not only a Load Balancer but it ensures high availability of the infrastructure. According to \citep{Verma:2015:LCM:2741948.2741964}, Kubernetes supports high-availability applications with run-time features that minimize fault-recovery time, and scheduling policies that reduce the probability of correlated failures. 

For offering a real high-available service, not only the Jupyter servers are spawned in containers managed by Kubernetes, but also the JupyterHub server itself. 
Additionally, the database of Kubernetes (ETCD) is distributed and replicated through the nodes so that the infrastructure becomes resistant to the failure of multiple nodes as shown in Figure~\ref{kube1}.


\subsection{Distributed Filesystem}


\begin{figure}[h]
\centering\includegraphics[width=0.9\linewidth]{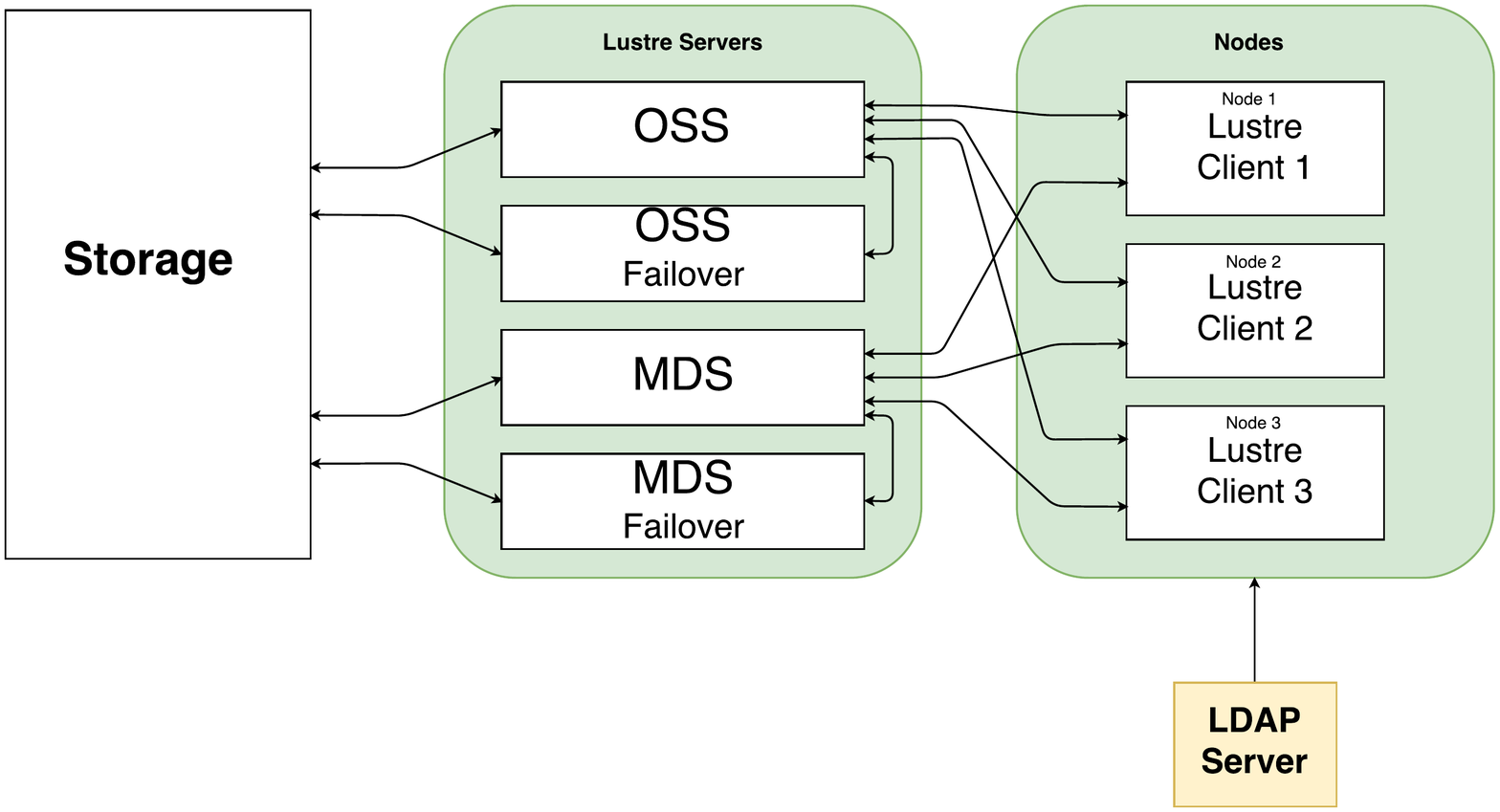}

\caption{Lustre infrastructure. The MGS is not shown since only manages the hosts. The failovers keeps the infrastructure working if the master server fails.}
\label{lustre}
\end{figure}

Another problem that arises from using multiple hosts is to provide distributed storage services. Moreover, current astronomical data analysis aims to process very large files that need to be accessed fast and often. Therefore, we need a filesystem that work through a high-speed network, with low latency and that can manage large file sizes. 

Lustre is a distributed filesystem optimized to operate with low latency at a TeraByte/PetaByte scale \citep{braam2002lustre}.
The main drawback of Lustre is that additional infrastructure is needed for its correct operation: not only an InfiniBand network is recommended, but also dedicated servers are required for attaining good performance. The Lustre infrastructure consist in three kind of servers: management servers (MGS), metadata servers (MDS) and object storage servers (OSS). When a client asks for a storage block, the petition is managed by the MGS, which looks up in the MDS for the block location and the user information (i.e., identification, privileges, etc). Finally the MGS establish a connection between the client and the OSS so the block can be mounted directly from the OSS. 

In our setup at the Chilean Virtual Observatory, we use $5$ servers for offering a high-available Lustre installation (i.e., 2 OSS, 2 MDS, and 1 MGS) as shown in Figure~\ref{lustre}. A dedicated InfiniBand network is used to connect Lustre severs and clients, and our two disk controllers are connected through optical fiber to the OSS servers. For JOVIAL we use Lustre to mount the users' files on the same nodes where Docker containers are hosted. Then, the mounted files can be passed to the containers as Docker volumes, so the whole Lustre architecture is completely hidden from the user. 

\subsection{JOVIAL Architecture}

\begin{figure}[ht]
\centering\includegraphics[width=0.8\linewidth]{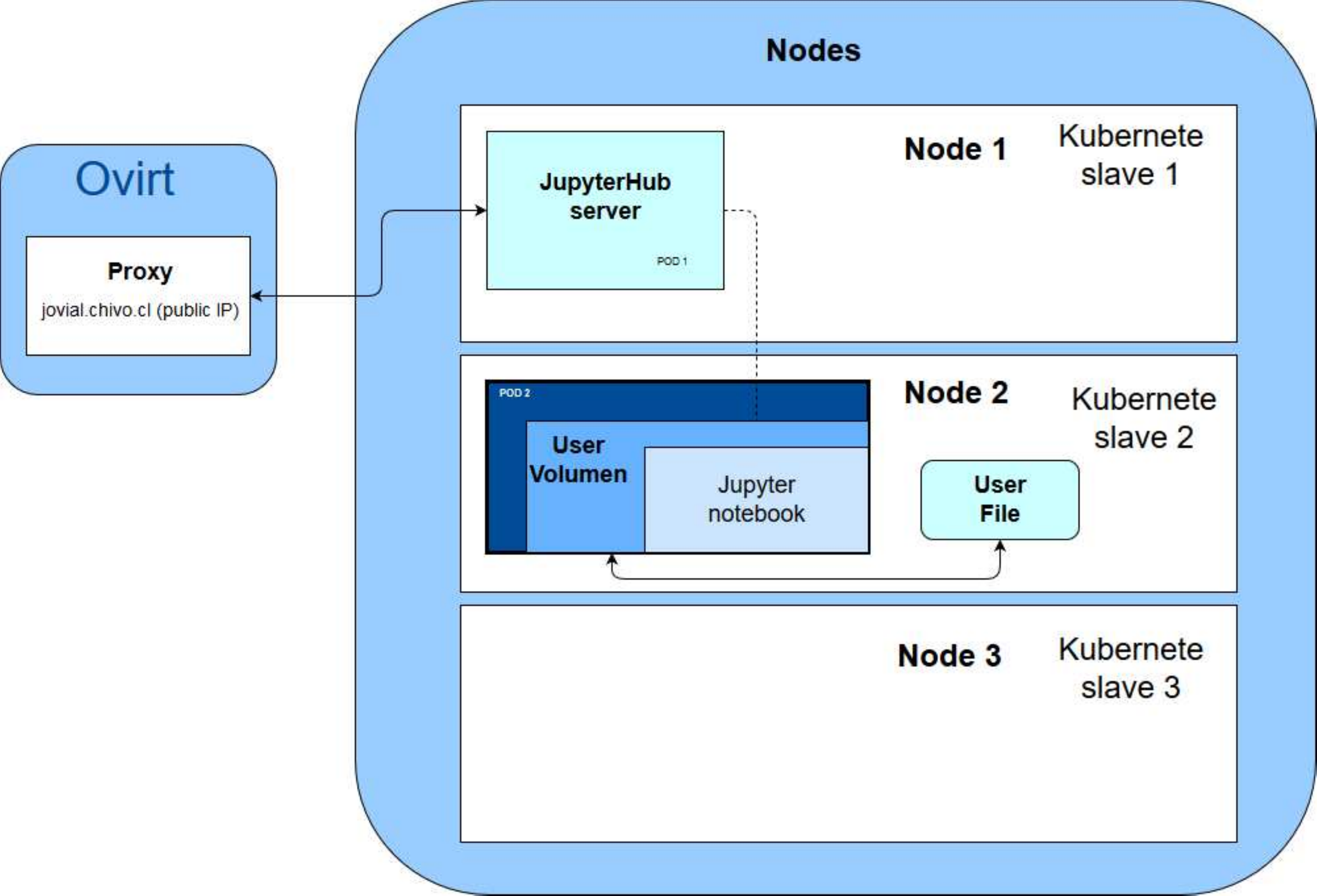}
\caption{JupyterHub-Kubernetes system. The JupyterHub server and the Jupyter Notebooks Servers run in Docker containers orchestrated by Kubernetes. User accounts are mounted through Lustre at the nodes and then are passed to the containers using Docker volumes and the KubeSpawner. All the nodes are in a private network and the JupyterHub servers can be accessed from the internet through a proxy.}
\label{kube2}
\end{figure}

We use $4$ computing nodes (servers) that work as hosts of the JupyterHub-Kubernetes system as shown in Figure~\ref{kube2}, which are also Lustre clients since they need to mount the users' files. 
Users are managed globally using LDAP, a software protocol that makes entries in which management data of the user is stored.
The Jupyter spawner that allows us to spawn notebook servers in Kubernetes clusters is called KubeSpawner. This module uses the global LDAP \texttt{id} and \texttt{username} to recreate the same user credentials in the container, and mounts the user home directory as a Docker volume. The recreated user owns the files mounted in the container since it has the same \texttt{id} and \texttt{username} as the original one. Finally, KubeSpawner runs the Jupyter Notebook server in an isolated but functional environment. We keep all the servers in a private network and the public access to the JupyterHub server is given by a proxy running on a virtual machine over oVirt. 

\begin{figure}[h]
\centering\includegraphics[width=1\linewidth]{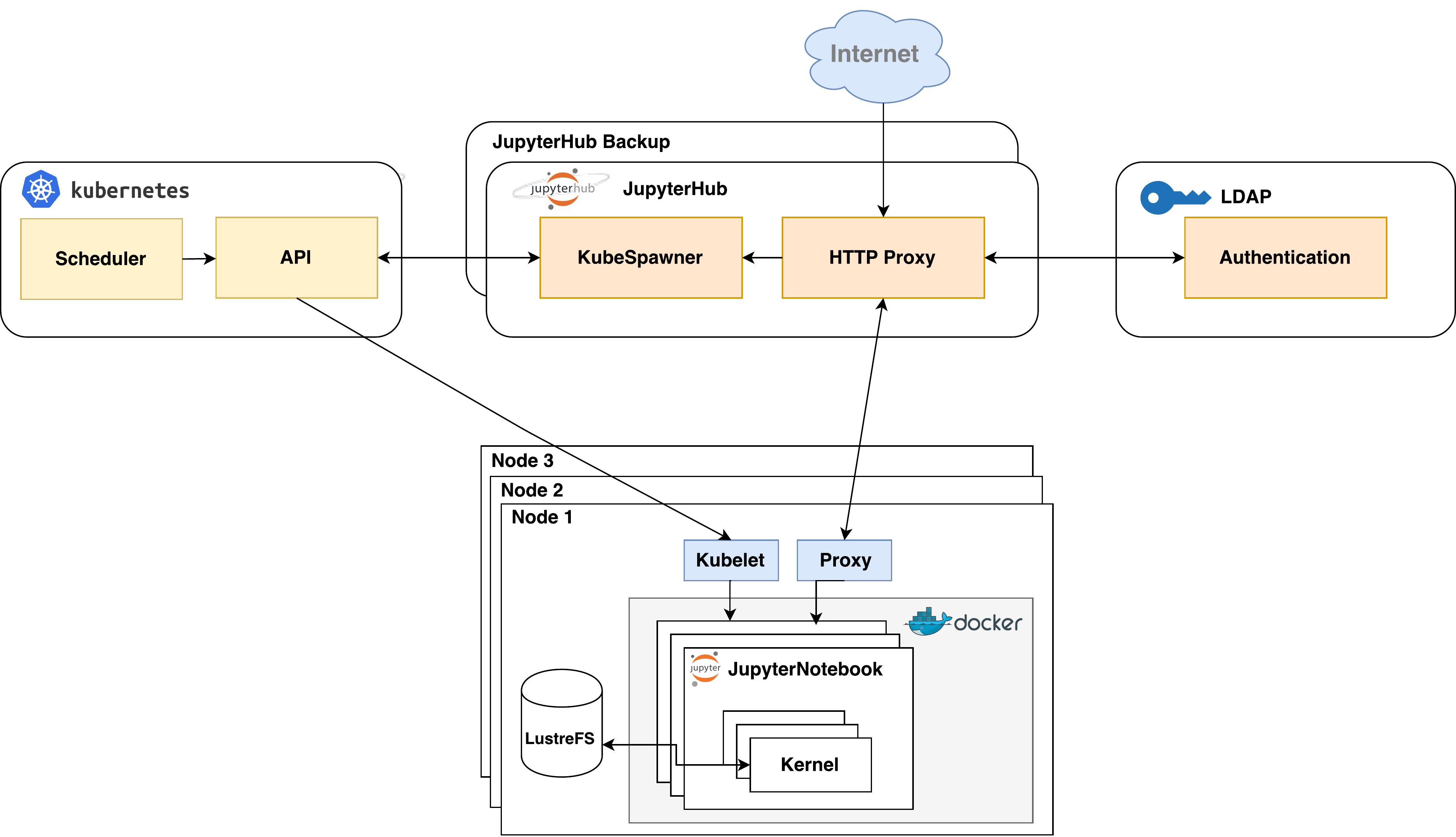}

\caption{JupyterHub is the responsible for the authentication and authorization using LDAP. Also, JupyterHub requests the creation of the notebook to Kubernetes. Kubernetes schedules a container on a node and the node starts the container using Docker. 
Lustre provide the user files to the container through a Docker volume. Moreover, there are multiple JupyterHub services running in case of a 
failure.}
\label{all}
\end{figure}

When a user enters the portal through the proxy and logs in into his account, JupyterHub checks if the container with the users' notebook server exists. If not, it asks Kubernetes to create it, spawn the files of the user into the container and contact the server. An overview of the components that interact in this process is shown in Figure~\ref{all}.



\section{Extending Jupyter's Astronomical Toolkit}
\label{sec:toolkits}

As presented in Section \ref{sec:nada}, some software packages are notebook friendly, but most of the astronomical data analysis software needs to be adapted for running into notebooks.
Porting legacy software to make it work over new systems is convenient for many reasons: usually is easier than re-implementing, reduces the risk of producing inconsistent software, and most important, it gives the possibility to make this software interact with new technologies such as Jupyter. 

The first barrier is binary and language compatibility. For example, the CUPID module of the Starlink suite \citep{berry2007cupid}, is written in C and uses data structures strongly rooted in Starlink libraries.
As CUPID does not use all the libraries, we developed wrappers using Cython, that uses only a few of the Starlink dynamic libraries. In this way it is possible to offer Python bindings, allowing the use of CUPID algorithms directly from the notebooks \citep{villanueva17adass}. This package is called \texttt{pyCupid}.

A second barrier are versions and compatibility. For example, even though CASA has Python bindings, and uses libraries that currently are notebook friendly (e.g., MatplotLib), the versions shipped by CASA are not. We have followed the \texttt{patchelf} strategy of CASANOVA, re-linking the binaries from the (old) self-provided
packages to the versions installed in JOVIAL. 
This allows compatibility with notebooks of both the CASA core and the CASA tasks \citep{mcmullin2007casa}. This package is called \texttt{Caspyter}.

\begin{figure}[h]
\centering\includegraphics[width=0.95\linewidth]{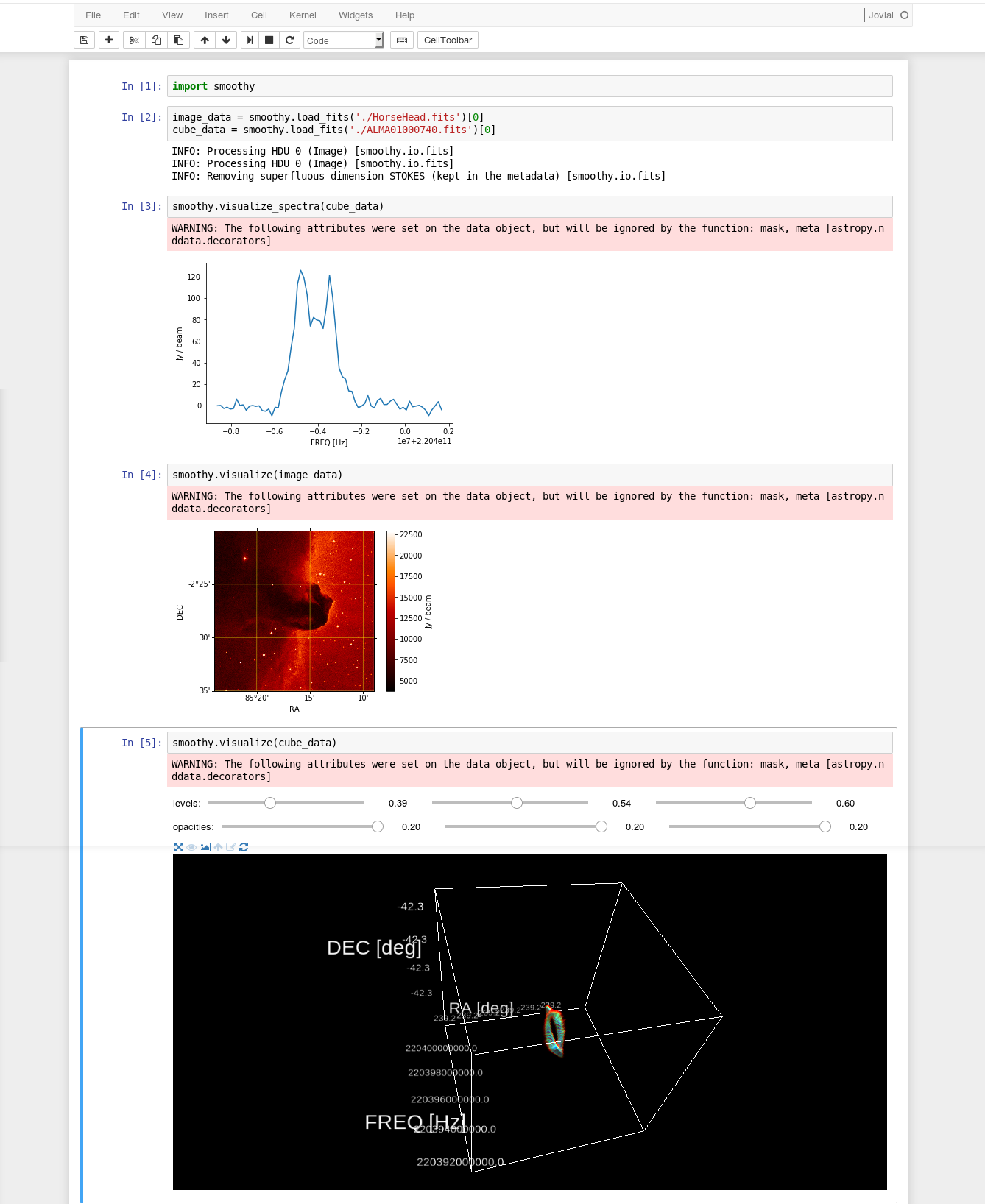}
\caption{JOVIAL running \texttt{Smoothy}. The example shows an image (optical) and a spectral cube (millimeter) both displayed using the function \texttt{visualize}. The spectra of the cube is also shown.}
\label{smoothy}
\end{figure}

The third barrier is package flooding. While having a large number of software packages for astronomy data analysis gives more options and functionality, it is not easy to handle the API details of each package. For this issue, we are developing a package that provides a unified object-oriented API for common tasks. This is implemented through suitable wrappers for data input and output operations provided by notebook-friendly packages. For example, the 
method \texttt{visualize(data)} will do its best to display the data. If it is an spectra, then will use \texttt{matplotlib}; if it is an image \texttt{AplPy} will be used; if it is a cube \texttt{PyVolume} will be used for a volumetric visualization. This package is called \texttt{smoothy}, and Figure \ref{smoothy} shows and example of its execution.

\section{Task Distribution}
\label{sec:distributed}

While Section \ref{sec:jovial} presents an architecture to scale in terms of users and notebooks, 
one of the main advantages of bringing code to data is task distribution across the data center infrastructure.  

We developed a prototype of a distributed pipeline for notebooks for finding regions of interest using a fast algorithm called RoISE \citep{mendoza16ac} that is implemented in the \texttt{ACAlib} python package \citep{maray2015:_acalib}. The main objective of this pipeline is to process a large number of data products with different resolutions, signal-to-noise ratios, densities, morphologies, imaging parameters, among others \citep{Araya17adass}.

As we need to process thousands of FITS data cubes, we evaluated tools to run hundreds of cpu-intensive and memory-intensive tasks in parallel. Between alternatives such as MPI, IParallel, Apache Spark or Hadoop MapReduce, we selected a Python library called Dask \citep{rocklin2015dask} mainly because required only a few modification to the original code.

Dask provides two main components: a set of dynamic task scheduling systems that is internally optimized for interactive and non-interactive computational workloads, and Big Data collections like parallel arrays, dataframes and lists that extend common interfaces like \texttt{NumPy}, \texttt{Pandas}, or Python iterators to larger-than-memory or distributed environments. These parallel collections run on top of the dynamic task schedulers. 



\begin{figure}[h]
\centering\includegraphics[width=0.9\linewidth]{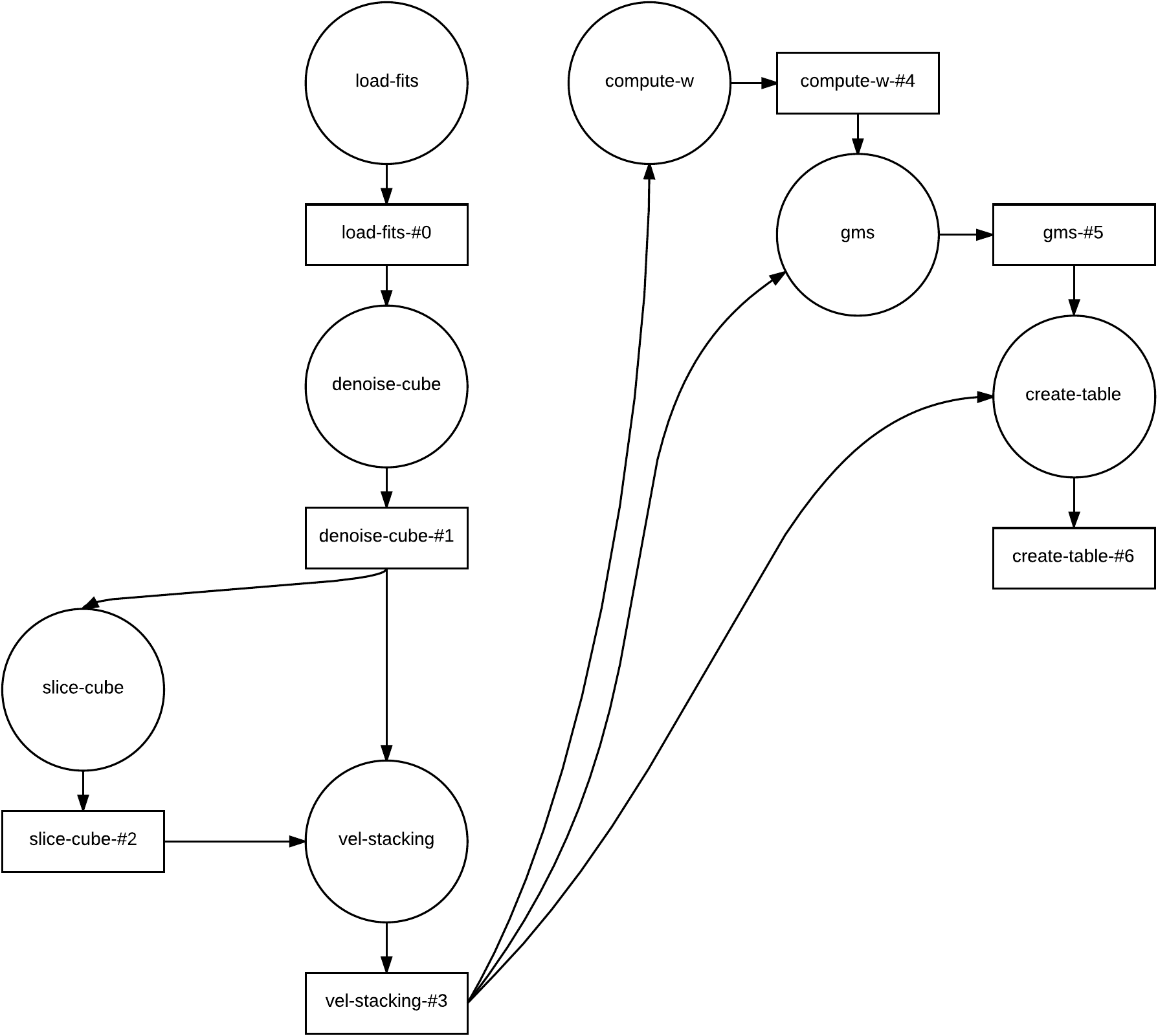}
\caption{Generated computational graph for the RoISE algorithm}
\label{roise_graph}
\end{figure}

In particular, the dask \emph{delayed} interface is particularly useful for parallelizing custom code. It is useful whenever the problem does not directly translate to a high-level parallel object like dask arrays or dataframes, but could still benefit from parallelism. It works by delaying the function evaluations and putting them into a dask graph.
We decided to use the \emph{delayed} object since the use of high-level objects like arrays or dataframes would require redesigning the whole algorithms. The \emph{delayed} object wraps the function calls into a lazy evaluated task graph so it can be later executed by one of the schedulers. Figure \ref{roise_graph} shows the version of the RoISE algorithm using the dask \emph{delayed} object; this pipeline represents all the operations that need to be executed to analyze one FITS cube. To process many of them, we used the dask distributed scheduler, which is backed by a single central scheduler process and many workers processes (i.e., python interpreters) spread across multiple machines.




We consider the dataset of all the FITS files produced by the ALMA pipeline, which are synthesized images and cubes from interferometric millimeter/sub-millimeter data. 
From these files we selected only the data cubes at the highest calibration level (i.e, primary-beam-corrected data with more that one spectral channel), which represents a total of 1.87 Terabytes stored in 4474 files. As users of JOVIAL, we have direct (read-only) access to the Chilean Virtual Observatory archive which is mounted within the nodes. 
As we needed to process hundreds of them in parallel, we moved the FITS files to Lustre in batches, allowing low latency and concurrent access. We were able to run the pipeline on 96 cores and using a total of 504 gigabytes of RAM, processing a total of 2666 ALMA data products in a few hours.




\begin{figure}[ht]
\centering\includegraphics[width=0.9\linewidth]{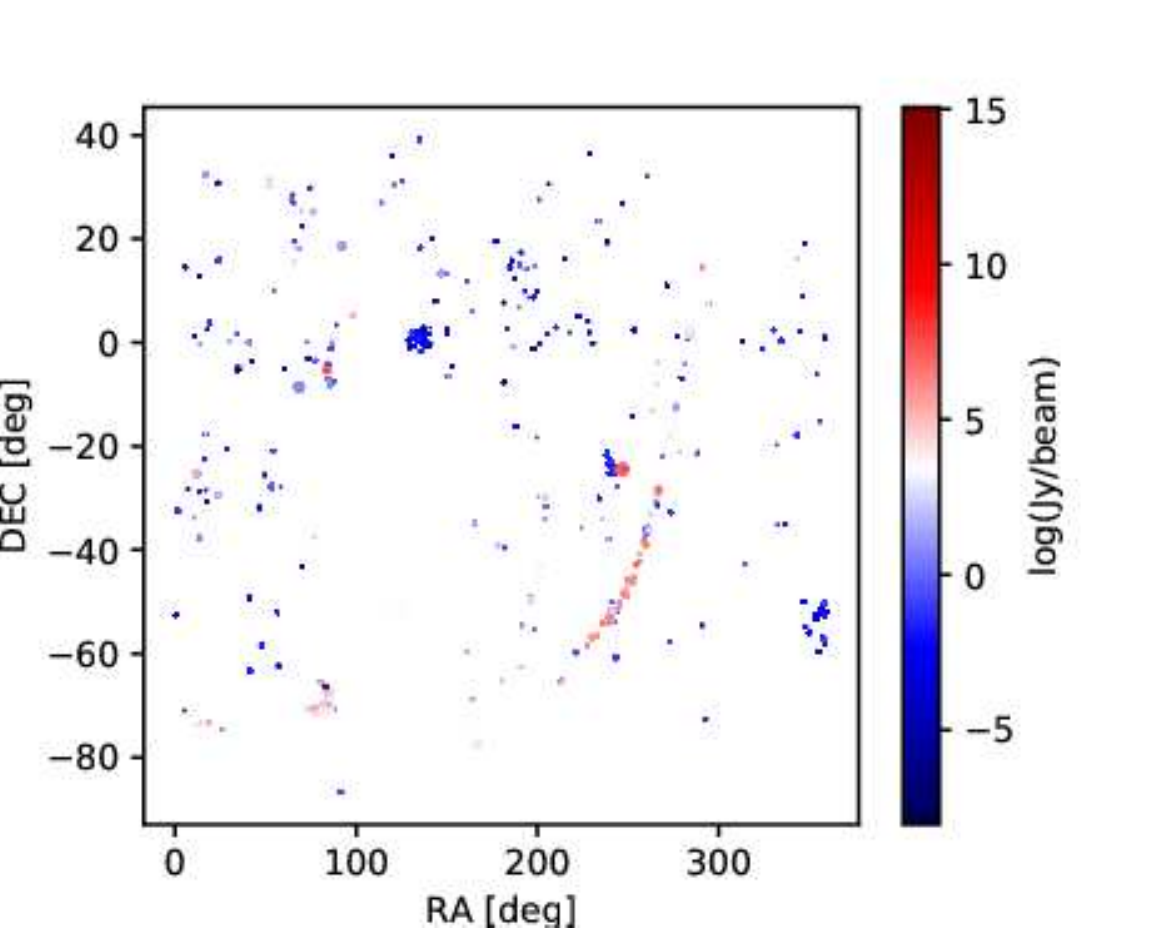}
\caption{Graphical summary of the intensities of the RoIs in sky coordinates. The size of each point also represents the area covered by the region.}
\label{g1}
\end{figure}

\begin{figure}[ht]
\centering\includegraphics[width=0.9\linewidth]{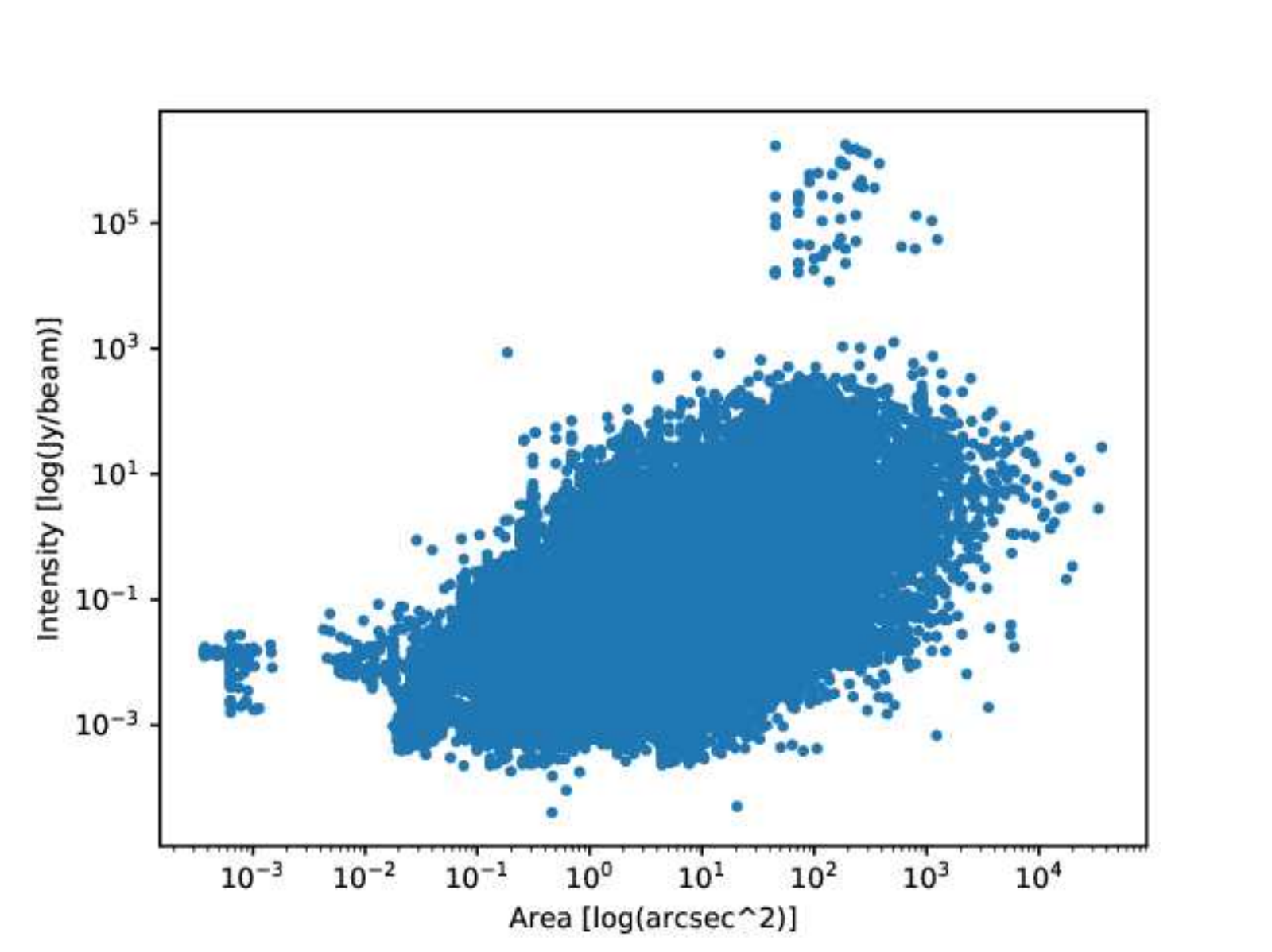}
\caption{Area vs Intensity in logarithmic scale for all RoIs. Three clusters are observed, where the ones with large area show low intensity and those with high intensity have a relatively small area.}
\label{g2}
\end{figure}

We found 47765 RoIs in all these files, for which we computed their centroid, semi-major/semi-minor axes, area, eccentricity, solidity, intensity and frequency range. As an example, we show in Figures \ref{g1} and \ref{g2} the summary of the results for intensity and area.

\section{Conclusions}
\label{sec:conclusions}

We have introduced JOVIAL, a multi-user data analysis service for astronomers in the Cloud that offers safe and secure Jupyter notebooks. The service is deployed under the high-availability and user-isolation principles, while maintaining all the flexibility and customization opportunities of Jupyter. We integrated several technologies such as JupyterHub, Docker, Kubernetes and Lustre to build the service, and the main configurations are available at its development web page \cite{jovial2018page}. We have equipped JOVIAL with popular Python libraries for astronomical data analysis, and developed a few others to fill functionality gaps with respect to popular desktop-based software. We showed that legacy or outdated astronomical data analysis software can be adapted to run in notebooks, and that the complexity of having too many libraries can be mitigated by integrated APIs. At last, we presented our prototype for distributing tasks across the data center, showing that modular code can be easily adapted to run tasks in parallel, for example, by using the delayed API of Dask. 

The main argument to build this service was to move the analysis near the data. The benefits of this are shown in Section~\ref{sec:distributed}, where a large dataset of FITS were analyzed from a robust Jupyter notebooks service, without transferring them through internet and using the high-performance resources of the data center.  

\subsection{Future Work}

In the short term, we have several undergoing projects to improve the user experience, the completeness and the efficiency of JOVIAL:
\begin{itemize}
    \item We are deploying an automatic user creation system for authorized domains, so astronomers can register to JOVIAL using their institutional e-mail.
    \item We are developing a bi-directional SAMP interface for Jupyter notebooks, so results can be transparently sent to (or received from) local desktop applications.
    \item Even though we have successfully integrated CASA to Jupyter notebooks, some GUIs are not suitable for the notebook concept. We are exploring IPython widgets to create panels that can replace this functionality.
    \item The only algorithm that was parallelized so far is RoISE, but there are several other algorithms developed by the Chilean Virtual Observatory that can benefit of Dask's delayed API. We plan to include parallel versions of them in the \texttt{ACALib} package, and include popular algorithms from other packages to be used within JOVIAL.
    \item At last, we plan to continue wrapping libraries into \texttt{Smoothy} in order to produce simple yet powerful notebooks under a homogeneous API.
\end{itemize} 

In the long term, we need to explore the possibility of providing collaborative notebooks (or at least shared notebooks) to promote the use of JOVIAL between groups of researchers, similar to what Google Docs is doing. In a similar direction, we need also to provide a proper notebook publishing mechanism, in order to encourage reproducible research through notebooks, and increase the visibility of the service.

\section{Acknowledgements}
This work has been partially funded by FONDEF IT 15I10041, CONICYT PIA/Basal FB0821 and CONICYT PIA/Basal FB0008. \newline

\label{sec:ack}

\section*{References}

\bibliography{mybibfile}

\end{document}